# Avancée-1 Mission and SaDoD Method:

## LiDAR-based stimulated atomic disintegration of space debris (SaDoD) using Optical Neural Networks

*Manuel Ntumba* [1], *Saurabh Gore* [2]

[1][2] *Division of Space Applications, Tod'Aérs - Transparent Organization for Aeronautics and Space Research, Togo.*

[1] *Regional Partnership Manager, Space Generation Advisory Council, Vienna, Austria.*

[1] *Government Scholar of Space Studies, Republic of Togo.*

[2] *Research Scholar, Moscow Aviation Institute, Moscow, Russia.*

manuel.ntumba@spacegeneration.org

sdgore@mai.education

## ABSTRACT

The surface degradation of satellites in Low Earth Orbit (LEO) is affected by Atomic Oxygen (AO) and varies depending on the spacecraft orbital parameters. Atomic oxygen initiates several chemical and physical reactions with materials and produces erosion and self-disintegration of the debris at high energy. This paper discusses Avancée-1 Mission, LiDAR-based space debris removal using Optical Neural Networks (ONN) to optimize debris detection and mission accuracy. The SaDoD Method is a Stimulated Atomic Disintegration of Orbital Debris, which in this case has been achieved using LiDAR technology and Optical Neural Networks. We propose Optical Neural Network algorithms with a high ability of image detection and classification. The results show that orbital debris has a higher chance of disintegration when the laser beam is coming from Geostationary Orbit (GEO) satellites and in the presence of high solar activities. This paper proposes a LiDAR-based space debris removal method depending on the variation of atomic oxygen erosion with orbital parameters and solar energy levels. The results obtained show that orbital debris undergoes the most intense degradation at low altitudes and higher temperatures. The satellites in GEO use Optical Neural Network algorithms for object detection before sending the laser beams to achieve self-disintegration. The SaDoD Method can be implemented with other techniques, but especially for the Avancée-1 Mission, the SaDoD was implemented with LiDAR technologies and Optical Neural Network algorithms.

## 1. INTRODUCTION

LiDAR-based object detection and recognition of the road environment was proposed by Kun Zhou et al. [3], focusing on detecting and tracking objects and recognizing lane markings and road features. Object tracking is performed using a 2-stage Kalman filter, taking into account the stability of tracking and accelerated movement of objects. [1] Lidar reflection intensity data is also used for sidewalk detection using robust regression. Road markings are detected using a modified Otsu method distinguishing rough and shiny surfaces. [4] Road reflectors that indicate the limit of the lane are sometimes obscured for various reasons. Lidar measurements help identify the obstacle's spatial structure, which makes it possible to distinguish objects according to their size and estimate the impact of driving on them. [2] Lidar systems provide better range and a wide field of view which helps detect obstacles on curves. Merging the lidar measurement with different sensors makes the system robust and useful in real-time applications, as lidar-dependent systems cannot estimate dynamic information about the detected object. [2] In laser satellite telemetry (SLR), a global network of observation stations measures the round trip time of the flight of ultra-short pulses of light to satellites equipped with retroreflectors. This provides instantaneous distance measurements with millimeter precision that can be accumulated to accurately measure orbits and a host of important scientific data. The laser pulse can also be reflected off the surface of a satellite without a retroreflector, which is used to track space debris. [5] Its ability to measure changes in the Earth's gravity field over time and monitor the movement of the network of stations relative to the geocenter and the ability to monitor vertical movement and long-term climate change [6] was providing a reference system for postglacial rebound, plate tectonics, sea level and ice volume change [7] determining the temporal mass redistribution of the solid earth system, ocean, and atmosphere [8] determination of orientation parameters of the Earth, such as coordinates of the Earth's poles and variations in the length of day [9] determination of precise satellite orbits for artificial satellites with and without active devices on board [10] [11] monitoring the response of the atmosphere to seasonal variations in solar heating. [12]

## 2. MISSION ANALYSIS

Orbital debris in low earth orbit is exposed to an environment of a dynamic mixture of different factors that influence its design, such as atomic species at low concentrations, charged particles, vacuum, micrometeorites, radiation, and temperature extremes. The presence of atomic oxygen can erode and degrade the thermal, mechanical, and optical properties of exposed orbital debris. The degree of surface degradation, represented here by the depth of erosion, decreases with increasing altitude, meaning that the orbital debris experiences the most intense degradation at lower altitudes. The depth of erosion due to the Atomic Oxygen attack of exposed materials is assessed by calculating the total fluency accumulated during the spacecraft mission. Even though the flow of atomic oxygen does not erode the surface, oxidation of the surface can alter the thermal properties of the surface layer, which would mean that the thermal requirements of the craft could be compromised by the change that results in equilibrium temperatures.

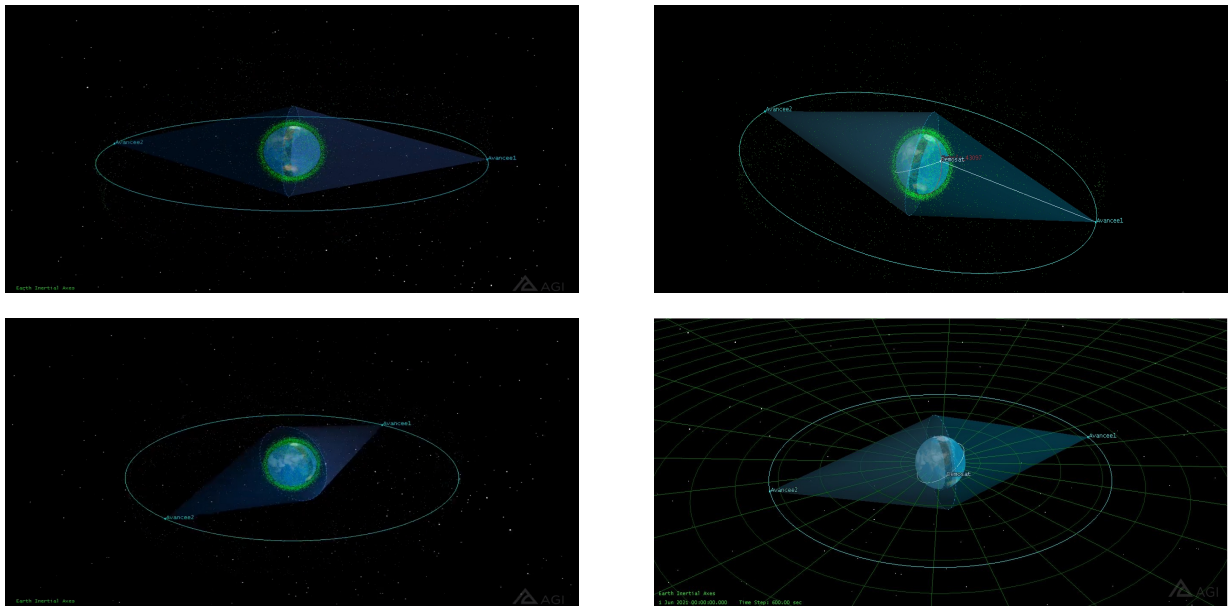

**Figure 1: Digital Mission Analysis of Avancée-1 Mission.**

3. **METHODS AND RESULTS**

Atomic oxygen present in a low Earth orbital (LEO) environment is formed by photo-dissociation of diatomic oxygen. The variation of the oxygen atomic erosion depth is a function of the spacecraft's orbital inclination, assuming a circular orbit, at mid-thermospheric altitude, for the mean level of solar activity. Atomic Oxygens are very corrosive, formerly combining with most of the materials they encounter. The AO collision with the orbital debris surfaces with this energy triggers many chemical and physical events. Surface erosion, loss of mass, degradation of mechanical, thermal, and optical properties, and changes in the chemical composition of materials can result from a collision with AO. Additionally, materials that resist the effects of AO erosion are likely to create an optical glow near the material's surface that can interfere with the visibility of optical systems of any given spatial structure. An additional factor that affects the degradation of the orbital debris is the orientation of the exposed surface relative to the attack of the atomic oxygen flow. The depth of erosion AO as a function of the orientation of the surface, taking polyimide as the target debris.

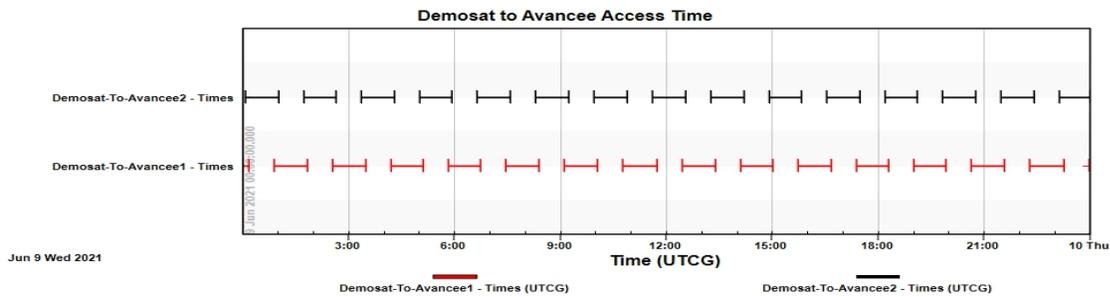

**Figure 2: Demosat access time of Avancée-1 Mission.**

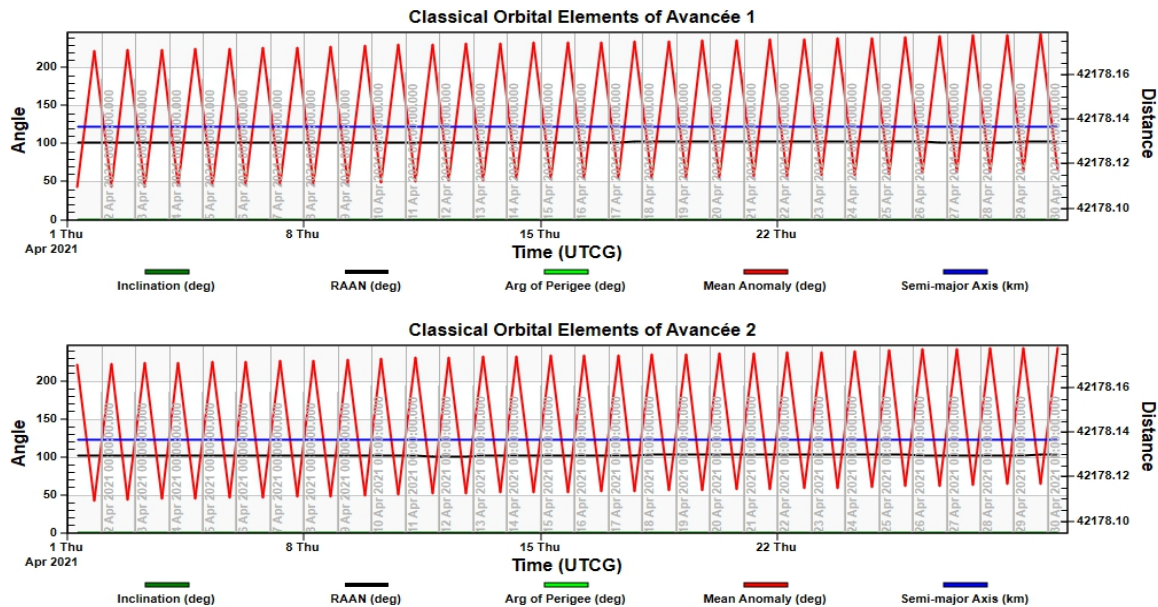

**Figure 3: Orbital Design of Avancée-1 Mission.**

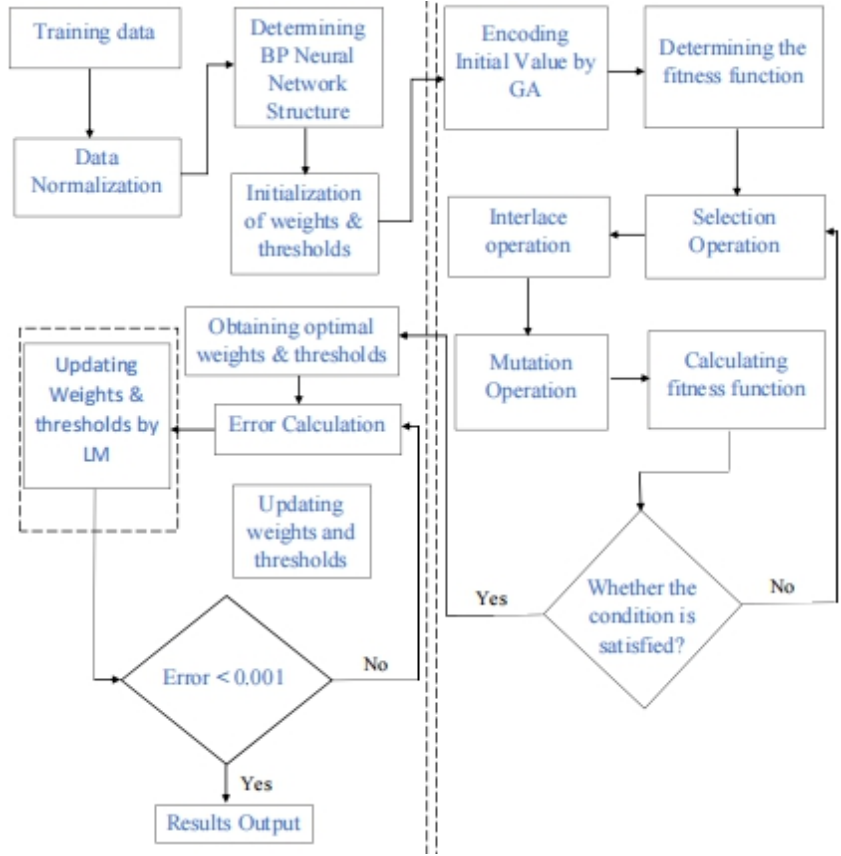

**Figure 4: Optical Neural Networks Model of Avancée-1 Mission.** An optical neural network algorithms are implemented with the optical components of the satellites. The optical configurations and optical models has many applications including image processing, pattern recognition , target tracking , real-time video processing.

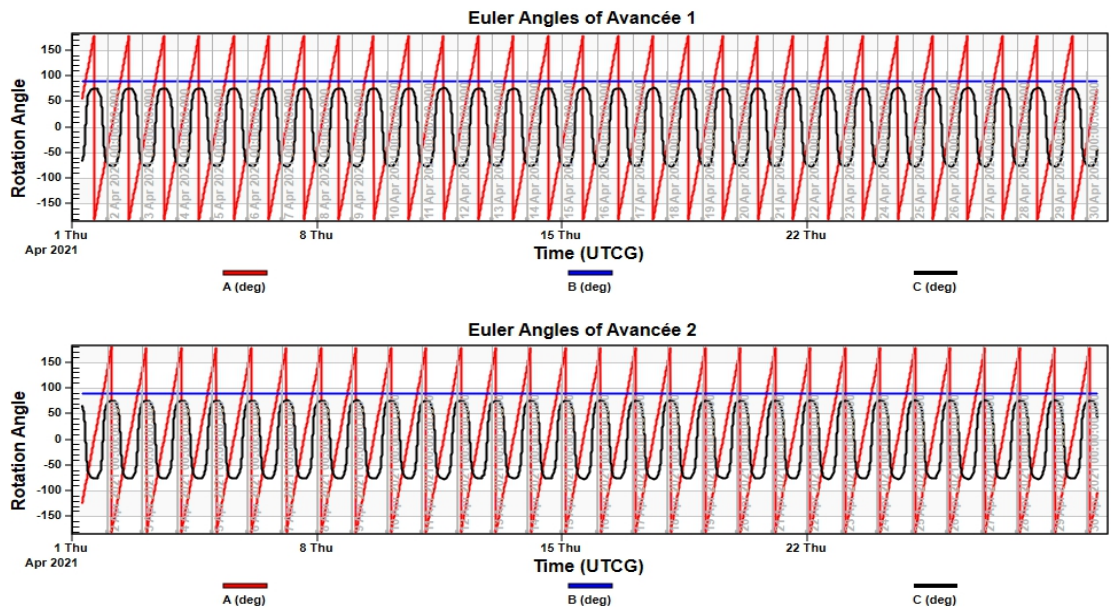

**Figure 5: Angular Analysis of Avancée-1 Mission.**

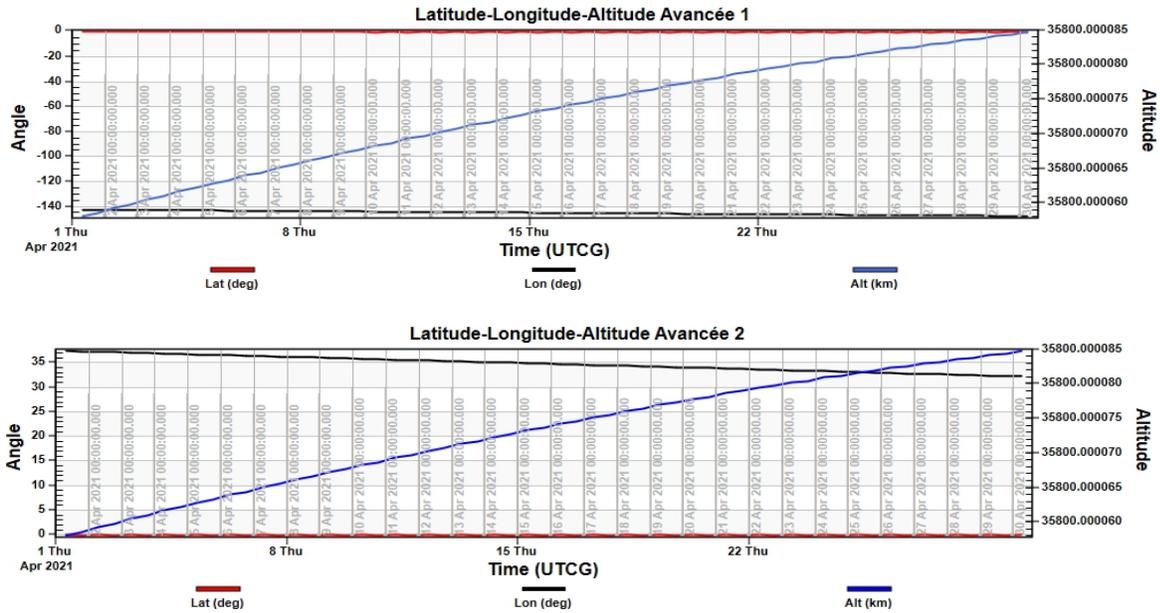

**Figure 6: Orbit Determination features of Avancée-1 Mission.**

4. **CONCLUSION**

In conclusion, Avancée-1 Mission launched during periods of minimal solar activity experiences less exposure to atomic oxygen than one launched during periods of high activity. Solar activity has a greater influence on the depth of erosion for higher altitudes. The solar radio frequency flux is an indicator of overall levels of solar activity as an indicator of extreme solar ultraviolet radiation. Effects of atomic oxygen in low earth orbits on Avancée-1 can be explained by the fact that solar activity significantly affects atmospheric density and Atomic Oxygen density. High solar activity results in high density, which is strongly affected by the variation in solar activity. The average Atomic Oxygen erosion depth varies due to consistent solar activity, which implies that the Sun emits more UV and X-rays during solar maximum. The orientation of the LiDAR is defined as a function of the attitude vector taken in this case as parallel to the speed vector and is considered in the frame of reference attached to the Avancée-1 satellite. The depth of erosion is greatest at the surface of the cylinder and decreases as the angle of orientation increases. In addition, the surface degradation behavior of orbital debris is variable due to atomic oxygen for different mission lengths for the average level of solar activity. The degree of surface degradation by the erosion depth increases with increasing mission duration. The effect of atomic oxygen can degrade the orbital debris in the presence of LiDAR beams. The atomic density of oxygen varies from 109 to 108 atoms per $cm^3$ over an altitude range of 300 to 500 km under high solar activity conditions, with low solar activity, concentrations above these altitudes are reduced to 108-106 atoms per $cm^3$.